\documentclass[letterpaper, 10 pt, conference]{ieeeconf}

\IEEEoverridecommandlockouts
\usepackage{amsmath}
\usepackage{amssymb}
\usepackage{graphicx}
\usepackage{booktabs}
\usepackage{array}
\usepackage{url}
\usepackage{cite}
\usepackage[T1]{fontenc}

\graphicspath{{figures/}}

\newcommand{\unit}[1]{\ensuremath{\mathrm{#1}}}
\newcommand{\degK}{\unit{K}}

\newcommand{\plantbench}{\texttt{plantbench}}

\title{\LARGE \bf A Plantwide Simulation Library for Control-Oriented Design: Analytical and Data-driven Use Cases}

\author{Kyle Territo$^{1}$, Luis A. Briceno-Mena$^{2}$ and Jose A. Romagnoli$^{1}$
\thanks{$^{1}$K. Territo and J. A. Romagnoli are with the Cain Department of Chemical Engineering, Louisiana State University, Baton Rouge, LA 70803, USA {\tt\small jose@lsu.edu}}%
\thanks{$^{2}$L. A. Briceno-Mena is with Data and Intelligent Systems, Princeton University, Princeton, NJ 08544, USA}%
}

\begin{document}

\maketitle
\thispagestyle{empty}
\pagestyle{empty}
\setcounter{footnote}{2}

\begin{abstract}
Control structures and operability indices for plantwide processes are developed and compared on fixed simulated flowsheets, and the configuration of a study is rarely reported in a form that another group can reproduce. In this contribution, \plantbench, a library of plantwide process simulations in which the design, the controller tuning and the instrumentation of a plant are declared as hashed plain-data configurations, is presented. On a heat-integrated reactor--separator--recycle plant with an open-loop unstable reactor, a heuristic plantwide control design prices four heat-exchanger networks (HENs) by the least damping ratio of the linearized closed loop, and 2016 closed-loop runs over networks, tuning, analyzer deadtime and upsets test that ranking. Results show that the damping ratio orders the networks as the runs do with the network that recovers the most heat reaching $0.013$ against $0.157$ for the selected one and cycling under a throughput increase, but that it is a weak detector of which operation will oscillate. At the recall of a random forest on the first $100\ \unit{min}$ after an upset, 60 of 82 limit cycles, a damping-ratio threshold raises 173 false alarms against 5. The integral time of the temperature loop, to which the damping ratio is nearly insensitive, decides in part whether the plant holds an operating point.
\end{abstract}

\section{Introduction}

Plantwide control design decides which variables are controlled, which inputs are paired with them and where production is set, for a plant whose units are coupled by material recycle and energy integration \cite{luyben1998,skogestad2000}. Its methods are developed on a small number of simulated test problems. The Williams--Otto plant \cite{williams1960}, the hydrodealkylation of toluene \cite{araujo2007}, the vinyl acetate monomer process \cite{luyben1998vam} and reactor--separator--recycle systems \cite{larsson2003,bildea2003} have served the design of control structures, and the Tennessee Eastman process \cite{downs1993,bathelt2015,reinartz2021} is a common reference for control and fault detection. Wastewater treatment offers the closest precedent for a benchmark with a stated evaluation protocol, fixing a plant, its disturbances, its sensors and its scoring indices \cite{jeppsson2007}. Learning-based control, in contrast, is developed mostly on single-unit environments that omit the couplings of a plant \cite{shin2019,nian2020,bloor2024}.

Each plantwide test problem is a fixed flowsheet. It offers the designs and disturbances its authors chose, rather than a design space over which a control method can be evaluated across alternatives, and reviews of the integration of design and control name the absence of standard test problems among the obstacles to routine practice \cite{ricardez2009,sharifzadeh2013,vega2014}. The same limitation affects the operability indices that such a design relies on. An index read from the closed loop linearized at the design point, such as its least damping ratio, is computed without a single run. Whether it predicts the operation that follows an upset, when valves saturate and integrators wind up, is an empirical question, and answering it requires many closed-loop runs across a design space, generated and reported so that another group can repeat them.

In this contribution, \plantbench\footnote{\url{https://github.com/lbrice1/plantbench}}, a library of plantwide process simulations, is presented, and used to compare two design approaches as a demonstration. We declare every choice that selects a plant from a design space, the HEN, the tuning of a loop and the deadtime of an analyzer, as a plain-data configuration that is hashed to identify its run and expanded from a specification into a reproducible dataset, and a user's own plant is brought under the same protocol through a case interface checked by a contract shipped with the library. The case study is a heat-integrated reactor--separator--recycle plant whose continuous stirred-tank reactor (CSTR) operates on the open-loop unstable branch of its multiplicity \cite{romagnoli2020}. We first design its control structure by the plantwide procedure of Luyben et al.~\cite{luyben1998} and price four HENs by the least damping ratio of the linearized closed loop. We then generate 2016 closed-loop runs over the networks, the tuning, the analyzer deadtime and six upsets, and test the damping ratio against the realized outcome in three ways: as a detector of limit cycles, against a classifier on the early response evaluated on folds that hold out whole combinations of plant and upset, and by variance-based sensitivity indices \cite{saltelli2010} over the design variables.

\section{Library and Protocol}
\label{sec:library}

\subsection{Cases and configurations}

A case of the library is a plant model with a steady-state design point, a verified reference solution and a design space. A configuration selects one member of that space through nine sections: the control structure, the case's options, parameter overrides by path, set-points, set-point changes, controller tuning, measurement non-idealities, valve non-idealities and a disturbance. A configuration is plain data; it round-trips exactly through \texttt{TOML}, a hash of it identifies its run, and every section is checked against the case when the plant is built, so that a misspelled loop or parameter raises an error.

A case supplies its design, control structures, disturbances and measurements through one interface, and the library applies configurations, datasets, tasks and sensitivity analysis to any case that implements it. A user develops a case in a package of their own, started from a template that the library generates, and declares it through a Python entry point, or registers it for a session with a decorator; the library then finds the case by its identifier in every process, including the parallel workers of data generation. A contract of checks shipped with the library, that the design point is a fixed point of every control structure to $10^{-9}$, that an undisturbed run stays at rest and that ideal instruments leave the closed-loop spectrum unchanged, is run on a user's case as on the library's own, and the manifest of a dataset over such a case records the version and commit of the package that defined it. A case that passes the contract and the library's rules on imports, documentation and tests is added to the library by moving its package into it.

A control structure is a list of proportional--integral (PI) loops and ratio stations declared as data. A primary loop writes the set-point of a secondary loop, the integral of each loop is appended to the plant state, and windup is handled by back-calculation \cite{astrom2006}, so that the closed loop is one system of ordinary differential equations with a continuous right-hand side. A measurement may carry a first-order lag and a deadtime represented by a Pad\'e approximation or a chain of first-order lags, and a valve a first-order travel and a rate limit; the closed loop therefore stays linearizable. Each disturbance declares the times at which it acts discontinuously, and the closed loop is integrated in segments between them by an automatic stiff/non-stiff method \cite{petzold1983} at relative and absolute tolerances of $10^{-8}$ and $10^{-10}$. The closed loop is linearized at the design point by central differences, and its least damping ratio $\zeta = -\mathrm{Re}\,\lambda/|\lambda|$ is taken over the modes with $|\mathrm{Im}\,\lambda| > 0.05\ \unit{min^{-1}}$.

\subsection{Specifications, tasks and verification}

A dataset is described by a specification that names a case, a seed and a base configuration, merges groups of partial configurations as a Cartesian product and draws a Latin hypercube over continuous configuration paths for every combination. The expansion is deterministic in the seed, and the dataset records a manifest of the specification, the library version and commit, and whether the source tree was clean. A task attached to the specification states the signals and window a method may see, the labeling rule, the folds and the metrics, and a digest of specification and task identifies the protocol, so that a second method reported against the same digest has been scored on the same folds. A digest is the first 12 hexadecimal characters of the SHA-256 of the canonical JSON of the specification, without its task for the runs (\texttt{pb-spec}) and with it for the protocol (\texttt{pb-protocol}), and \texttt{plantbench verify} recomputes both from the specification file. Table~\ref{tab:card} reports the dataset of this paper in that form. The case study reproduces the reference values of its reactor's steady state and eigenvalues ($+0.0600$, $-0.1350$ and $-0.7990\ \unit{min^{-1}}$) and its reference results byte for byte under the library's test suite of 329 tests \cite{plantbench}.

\begin{table}[t]
  \caption{The enumerated dataset as the library reports it}
  \label{tab:card}
  \centering
  \footnotesize
  \begin{tabular}{@{}l>{\raggedright\arraybackslash}p{0.68\columnwidth}@{}}
    \toprule
    Entry & Value \\
    \midrule
    Case & \texttt{reactor\_separator\_recycle}; study \texttt{regimes} \\
    Code & \plantbench{} 0.1.0 \\
    Environment & Python 3.14.7, NumPy 2.5.3, SciPy 1.18.1 \\
    Varied & D1, D3, D5; D4 at pump-around fractions $0$, $0.3$, $0.5$, $0.7$; six upsets at $30\ \unit{min}$ \\
    Sampled & 48 Latin hypercube points per combination: $K_c \in [-2, -0.5]\ \unit{m^3/(min\,K)}$ and $\tau_I \in [5, 40]\ \unit{min}$ of the reactor temperature loop, analyzer deadtime in $[0, 10]\ \unit{min}$; seed 20260918 \\
    Runs & $7 \times 6 \times 48 = 2016$, all completed; $1500\ \unit{min}$ at $2\ \unit{min}$ spacing; $6.89\ \unit{h}$ of integration \\
    Label & late outcome over the last $300\ \unit{min}$ \\
    Task & 16 signals over $0$ to $130\ \unit{min}$; five stratified folds, seed 0; balanced accuracy, limit cycles found, false alarms \\
    Specification & \texttt{studies/regimes/spec.toml}; \texttt{pb-spec:6a6fc673261b} \\
    Protocol & specification with its task; \texttt{pb-protocol:dab0d47475e9} \\
    \bottomrule
  \end{tabular}
\end{table}

\section{Case Study and Design Problem}
\label{sec:case}

\begin{figure*}[t]
  \centering
  \includegraphics[width=\textwidth]{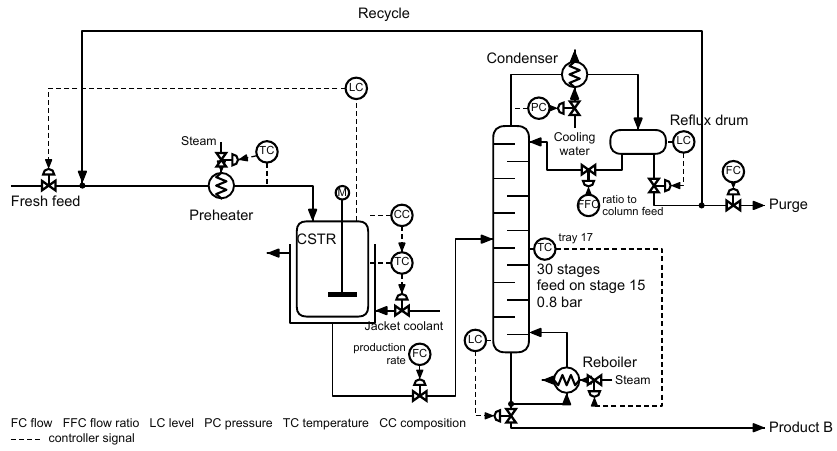}
  \caption{Reference control structure of the case study.}
  \label{fig:structure}
\end{figure*}

A fresh feed of reactant A, carrying $0.5\ \unit{mol\%}$ of a light inert, joins a recycle and enters a jacketed CSTR, in which the first-order exothermic reaction $\mathrm{A} \rightarrow \mathrm{B}$ runs on the open-loop unstable middle branch of the reactor's multiplicity. The effluent is separated in a thirty-stage column; the inert and the unconverted reactant leave overhead and are split into the recycle and a purge. The base plant has 134 states. Its design point is fixed by the plantwide balances together with three specifications that the balances leave free, the inert level at the reactor inlet ($0.549\ \unit{kmol/m^3}$) and the holdups of the reflux drum ($12\ \unit{kmol}$) and the column base ($6.90\ \unit{kmol}$). Its reference control structure (Fig.~\ref{fig:structure}) fixes the reactor effluent flow, holds the reactor level on the fresh feed, controls the tray~17 temperature on the boil-up, and moves the reactor temperature set-point from a composition loop.

Four HENs can be placed on the plant. D1 recovers no heat but carries the feed-temperature loop of the integrated designs; D3 is a feed--effluent heat exchanger at $0.4\ \unit{bar}$; D4 is a vacuum column at $0.15\ \unit{bar}$ whose reboiler takes the reactor effluent and a pump-around of reactor heat; and D5 sends the condenser duty to the reactor feed. The pump-around fraction of D4 is the fraction of the reactor's heat duty that the pump-around carries to the reboiler instead of the jacket. The design problem is to choose the network, the pump-around fraction, the tuning of the reactor temperature loop and the analyzer that maximize operating profit and minimize the global warming potential (GWP) of the energy the plant uses, while the plant stays reliable under six upsets: throughput changes of $+10\%$, $+20\%$ and $-10\%$, a $50\%$ increase in the inert content of the feed, a feed $10\ \unit{K}$ colder, and a step of $-0.05\ \unit{kmol/m^3}$ in the composition set-point.

The outcome of a run is read from its reactor temperature over the last $300\ \unit{min}$. A limit cycle swings by more than $0.5\ \unit{K}$ peak to peak, with the swing over the last $150\ \unit{min}$ at least half that over the $150\ \unit{min}$ before; an unsettled run swings by more than $0.05\ \unit{K}$ without sustaining it; a saturated run has a loop's input on a limit over its last $30\ \unit{min}$; and a settled run is none of these.

\section{Analytical Plantwide Design}
\label{sec:design}

The procedure of Luyben et al.~\cite{luyben1998} fixes the structure step by step on the model. The reactor temperature loop is tuned for damping rather than stability alone: at $K_c = -2\ \unit{m^3/(min\,K)}$ and $\tau_I = 5\ \unit{min}$ the least damping ratio is $0.061$, and at $K_c = -1$ and $\tau_I = 20\ \unit{min}$ it is $0.158$. Fixing a flow in the recycle loop is necessary and not sufficient. Under a $10\%$ increase in fresh feed the recycle grows by $5.22$ times the production increase with no flow fixed, by $4.51$ times with the reactor effluent fixed, and by $1.46$ times once the composition cascade holds the conversion per pass.

\begin{table}[t]
  \caption{Networks at the design point, and D4 over its pump-around fraction}
  \label{tab:designs}
  \centering
  \footnotesize
  \setlength{\tabcolsep}{4pt}
  \begin{tabular}{@{}lrrrrr@{}}
    \toprule
    Design & $P$ & Recovered & Utilities & Energy GWP & $\zeta$ \\
     & bar & GJ/h & \$/h & t/h & \\
    \midrule
    D1 & $0.80$ & $0$    & $477$ & $5.099$ & $0.155$ \\
    D3 & $0.40$ & $26.5$ & $381$ & $4.049$ & $0.157$ \\
    D5 & $0.80$ & $14.8$ & $388$ & $4.121$ & $0.155$ \\
    D4, fraction $0.7$ & $0.15$ & $67.7$ & $516$ & $3.700$ & $0.013$ \\
    D4, fraction $0.5$ & $0.15$ & --- & $533$ & $3.886$ & $0.048$ \\
    D4, fraction $0.3$ & $0.15$ & --- & $550$ & $4.072$ & $0.087$ \\
    D4, fraction $0$   & $0.15$ & --- & $576$ & $4.352$ & $0.159$ \\
    \bottomrule
  \end{tabular}
\end{table}

The networks separate on the environmental objective and on the damping ratio (Table~\ref{tab:designs}). At the design throughput the profits of the three integrated designs lie within $\$135/\unit{h}$ of one another. D4 recovers the most heat and reaches the lowest energy GWP, but its least damping ratio is $0.013$ against $0.155$ to $0.157$ for the others; all four share a rightmost eigenvalue of $-0.0028\ \unit{min^{-1}}$, so the distinction is damping rather than stability. The mechanism is the jacket, as the pump-around leaves it to run at $0.267\ \unit{m^3/min}$ against $1.092$ in the other designs, so that it has less room to close before the valve reaches its limit. The trade is monotonic in the pump-around fraction, D4 emits more than D3 at a fraction of $0.3$, at $4.072$ against $4.049\ \unit{t/h}$, and less at $0.5$, where its damping ratio has already fallen to $0.048$. Retuning does not recover it since over thirty settings of the temperature loop only eighteen stabilize D4, and the best damping ratio reached is $0.035$.

\begin{figure}[t]
  \centering
  \includegraphics{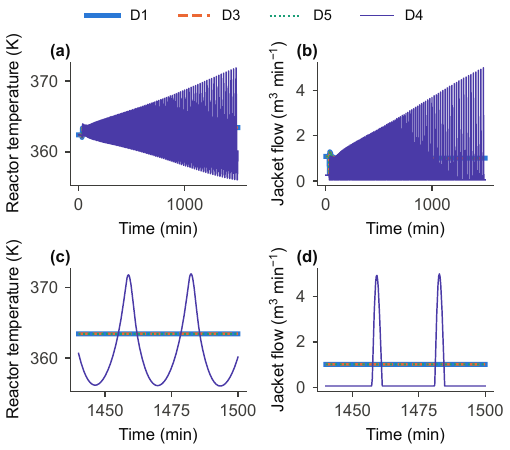}
  \caption{Response of the integrated designs to a $10\%$ throughput increase. (a) Reactor temperature. (b) Jacket coolant flow. (c, d) The same over the last $60\ \unit{min}$. D1, D3 and D5 coincide.}
  \label{fig:response}
\end{figure}

A $10\%$ throughput increase shows what the damping ratio measures (Fig.~\ref{fig:response}). D1, D3 and D5 settle, with an integral absolute error in reactor composition of $20.20\ \unit{kmol\,min/m^3}$. That is not the case for D4, its error is $291.24$, its jacket valve is on a limit for $75\%$ of the run, and its reactor temperature oscillates between $356.05$ and $371.87\ \degK$ with a period of $22.8\ \unit{min}$. Because the valve spends most of each period on a limit, the amplitude of the cycle is set by the valve limits and the anti-windup rather than by the linearization. The procedure selects D3, which reaches the minimum-utility target of pinch analysis \cite{linnhoff1983} at its pressure with one exchanger and is ahead of D5 on both objectives, and leaves D4 at a high pump-around fraction as a design known to cycle. It reads the damping ratio at the design point with every instrument ideal, so it does not say how the plant behaves across the tuning, the analyzer deadtime and the upsets.

\section{Closed-Loop Evaluation of the Damping Ratio}
\label{sec:data}

\subsection{The enumerated dataset}

The dataset of Table~\ref{tab:card} combines seven plants with six upsets and 48 sampled settings of the tuning and the analyzer deadtime. All 2016 runs completed, in $6.89\ \unit{h}$ of integration. Of these, $78.4\%$ settled, $17.5\%$ remained unsettled and $4.1\%$, 82 runs, ended in a limit cycle; none ended saturated. The unsettled runs are almost entirely those of the inert upset, which moves the plant on a time scale longer than the run. Limit cycles concentrate in D4 and grow with its pump-around fraction, from 3 of 288 runs at a fraction of $0$ to 53 of 288 at $0.7$, and 50 of the 82 follow the $20\%$ throughput increase.

\begin{figure}[t]
  \centering
  \includegraphics{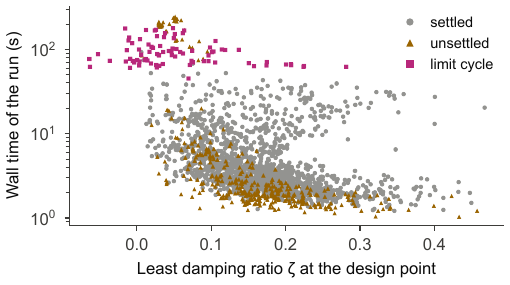}
  \caption{Wall time of each run of the enumerated dataset against the least damping ratio at its design point.}
  \label{fig:damping}
\end{figure}

The damping ratio separates the outcomes only in part (Fig.~\ref{fig:damping}). Its median is $0.036$ among the limit cycles against $0.166$ among the settled runs, and as a detector of limit cycles it reaches an area under the receiver operating characteristic curve (ROC AUC) of $0.872$. Limit cycles occur up to $\zeta = 0.281$, however, and 1803 runs at or below that value do not cycle. The cost of a run separates them more sharply as a limit cycle costs a median of $86.60\ \unit{s}$ to integrate against $3.52\ \unit{s}$ for a settled run.

\subsection{Recognizing a limit cycle early}

We ask whether the first $100\ \unit{min}$ after the upset predict the late outcome better than the design point does. Each run is one observation: its 16 signals, the reactor temperature, composition and volume, the drum and base levels, the tray~17 temperature, the reactant in the bottoms, the recycle flow, the reactor inlet temperature and the manipulated variables, are sampled every $10\ \unit{min}$ as means over $8\ \unit{min}$ and flattened into 176 features. We compare a ten-nearest-neighbor (10-NN) classifier and a random forest of 500 trees \cite{breiman2001} with class weights inversely proportional to the outcome counts, on the task's stratified folds and on folds grouped by plant and upset, which hold out whole combinations so that none of a test combination's runs is seen in training. The baselines are the damping ratio and rightmost eigenvalue of the design point, and the design variables themselves.

\begin{table}[t]
  \caption{Prediction of the late outcome on the task's folds and on folds grouped by plant and upset. BA: balanced accuracy; LC: limit cycles found of 82; FA: false alarms}
  \label{tab:early}
  \centering
  \footnotesize
  \setlength{\tabcolsep}{3pt}
  \begin{tabular}{@{}llrrrrrr@{}}
    \toprule
    & & \multicolumn{3}{c}{Task folds} & \multicolumn{3}{c}{Grouped folds} \\
    \cmidrule(lr){3-5}\cmidrule(l){6-8}
    Features & Classifier & BA & LC & FA & BA & LC & FA \\
    \midrule
    Early window & 10-NN  & $0.837$ & 48 & 3  & $0.754$ & 28 & 0 \\
    Early window & forest & $0.949$ & 73 & 5  & $0.884$ & 60 & 5 \\
    Design variables & forest & $0.888$ & 59 & 15 & $0.694$ & 14 & 15 \\
    $\zeta$, rightmost $\lambda$ & 10-NN & $0.439$ & 25 & 11 & $0.428$ & 24 & 20 \\
    \bottomrule
  \end{tabular}
\end{table}

The early response carries information that generalizes to unseen combinations of plant and upset, and the design point does not (Table~\ref{tab:early}). On grouped folds the forest on the early window keeps a balanced accuracy of $0.884$ and finds 60 of 82 limit cycles with 5 false alarms, while the forest on the design variables falls to $0.694$ and finds 14. At matched recall the damping ratio is a weak detector since to find the same 60 limit cycles a threshold on $\zeta$ must be set at $0.0744$, and it then raises 173 false alarms; to find the 28 of the 10-NN classifier it raises 17 against 0. The threshold is read on all runs, which favors the damping ratio. The classes are unbalanced, so the limit cycles found and the false alarms, rather than the balanced accuracy, are the figures that bear on the design question.

\subsection{Mechanism and sensitivity}

Reading the last $300\ \unit{min}$ of the runs by the data mining workflow reported elsewhere \cite{brice2022dce,seghers2023}, embedding by pairwise controlled manifold approximation and projection \cite{wang2021}, density-based clustering \cite{mcinnes2017} and subspace greedy search \cite{zhu2018} of the variables that separate two clusters, recovers the two paths of Section~\ref{sec:design} without a design variable or an outcome among the inputs. With the samples taken as deviations from the design point, the cluster holding the most limit-cycle samples differs from the largest cluster first through the reactor inlet temperature ($25.9\%$ of the contributions), the feed-temperature path; with the samples taken as deviations from each run's own late-window mean, the largest unsettled cluster differs through the jacket coolant flow ($67.7\%$).

\begin{table}[t]
  \caption{Total Sobol' indices over the design variables of D4, $n = 256$, at a $10\%$ throughput increase, with $95\%$ bootstrap intervals. Pump-around fraction in $[0, 0.7]$, column pressure in $[0.15, 0.275]\ \unit{bar}$, $K_c \in [-4, -0.25]\ \unit{m^3/(min\,K)}$, $\tau_I \in [5, 160]\ \unit{min}$, analyzer deadtime in $[0, 10]\ \unit{min}$}
  \label{tab:sobol}
  \centering
  \footnotesize
  \setlength{\tabcolsep}{3pt}
  \begin{tabular}{@{}lll@{}}
    \toprule
    Design variable & Damping ratio & Realized outcome \\
    \midrule
    Analyzer deadtime    & $0.480$ $[0.382, 0.596]$ & $0.736$ $[0.519, 1.046]$ \\
    Pump-around fraction & $0.244$ $[0.194, 0.305]$ & $0.584$ $[0.373, 0.842]$ \\
    $K_c$                & $0.599$ $[0.467, 0.760]$ & $0.330$ $[0.163, 0.536]$ \\
    $\tau_I$             & $0.025$ $[0.015, 0.037]$ & $0.305$ $[0.158, 0.504]$ \\
    Column pressure      & $0.000$ $[0.000, 0.001]$ & $0.000$ $[0.000, 0.000]$ \\
    \bottomrule
  \end{tabular}
\end{table}

The sensitivity indices compare what the criterion responds to with what decides the outcome (Table~\ref{tab:sobol}). Over a separate sample of D4 designs, with the pump-around fraction and the column pressure resolved continuously and the upset held at the $10\%$ throughput increase, the damping ratio responds most to the gain of the temperature loop and to the analyzer deadtime, and hardly to the integral time. The realized outcome, whether a run holds an operating point, responds to the deadtime, the pump-around fraction, the gain and the integral time. Its first-order indices are far below its totals, $0.203$ against $0.736$ for the deadtime, so the outcome is carried by interactions among the design variables. The rankings differ most in the integral time, at a total index of $0.025$ for the damping ratio and $0.305$ for the outcome.

\section{Conclusions}

A library of plantwide process simulations has been presented in which the design, the tuning and the instrumentation of a plant are plain-data configurations, and a study is reported through a specification and a task whose digests identify its runs and its protocol. On its heat-integrated case study, the damping ratio of the linearized closed loop ordered the networks as 2016 closed-loop runs did, rejecting the network that recovers the most heat at $0.013$ against $0.157$. As a detector of the runs that cycle it was weak, on folds that hold out whole combinations of plant and upset, a random forest on the first $100\ \unit{min}$ after the upset found 60 of 82 limit cycles with 5 false alarms, where a threshold on the damping ratio at the same recall raised 173.

Three limitations bound these results. They are model-to-model, with outcomes that the simulator defines. The analyzer deadtime enters through a second-order Pad\'e approximation, and at a pump-around fraction of $0.7$ the damping ratio of D4 is $0.0125$, $0.2863$ and $0.0664$ at deadtimes of $0$, $3$ and $10\ \unit{min}$, a non-monotonic dependence whose fidelity to a true delay we have not established. And the inert upset moves the plant on a time scale longer than the $1500\ \unit{min}$ horizon.

Whether the heat-integrated plant holds an operating point depends on the integral time of its temperature loop, to which the damping ratio at the design point is nearly insensitive. A plant whose reliability is decided by a limit cycle under saturation is therefore better screened on its closed-loop response than on its linearization alone.

\bibliographystyle{IEEEtran}
\bibliography{refs}

\end{document}